\documentclass{aa}
\usepackage{graphicx}
\usepackage{times}
\begin{document}

\def\kms{\mbox{km\,s$^{-1}$}}
\def\Hubble{\mbox{km\,s$^{-1}$\,Mpc$^{-1}$}}
\def\Doppler{\mathcal{D}}
\def\lsim{\raisebox{-.5ex}{$\;\stackrel{<}{\sim}\;$}}
\def\gsim{\raisebox{-.5ex}{$\;\stackrel{>}{\sim}\;$}}
\def\Snutspace{$(S,\nu,t)$-space}
\def\lgSnutspace{$(\lg S,\lg \nu,\lg t)$-space}
\newcommand{\mrm}[1]{\mathrm{#1}}
\newcommand{\dmrm}[1]{_{\mathrm{#1}}}
\newcommand{\umrm}[1]{^{\mathrm{#1}}}
\newcommand{\Frac}[2]{\left(\frac{#1}{#2}\right)}
\newcommand{\eqref}[1]{Eq.~(\ref{#1})}
\newcommand{\eqsref}[2]{Eqs~(\ref{#1}) and (\ref{#2})}
\newcommand{\eqssref}[2]{Eqs~(\ref{#1}) to (\ref{#2})}
\newcommand{\figref}[1]{Fig.~\ref{fig:#1}}
\newcommand{\tabref}[1]{Table~\ref{tab:#1}}
\newcommand{\secref}[1]{Sect.~\ref{sec:#1}}
\newcommand{\url}[1]{{\ttfamily\small #1}}

%\thesaurus{00(0;0;0)}
\title{The  INTEGRAL Science Data Centre (ISDC)}
\author{
T.J.-L. Courvoisier \inst{1,}\inst{2}, 
R. Walter \inst{1,}\inst{2},
V. Beckmann \inst{1,}\inst{3},
A.J. Dean \inst{4},
P. Dubath \inst{1,}\inst{2},
R. Hudec \inst{5},
P. Kretschmar \inst{1,}\inst{10},
S.~Mereghetti \inst{6},
T. Montmerle \inst{7},
N. Mowlavi \inst{1,}\inst{2},
S. Paltani \inst{8}
A. Preite Martinez \inst{9},
N. Produit \inst{1,}\inst{2},
R. Staubert \inst{3},
A.W.~Strong \inst{10},
J.-P. Swings \inst{11},
N.J. Westergaard \inst{12},
N. White \inst{13},
C. Winkler \inst{14},
A.A. Zdziarski \inst{15}
}
\institute{
{\it INTEGRAL} Science Data Centre, ch. d'\'Ecogia 16, CH-1290 Versoix, Switzerland \and
Geneva Observatory, ch. des Maillettes 51, CH-1290 Sauverny, Switzerland \and
Universit\"at T\"ubingen, Inst. f\"ur Astronomie und Astrophysik, Abt. Astronomie, Sand 1, D--72076 T\"ubingen, Germany \and
Astronomy Group, Physics Department,University of Southampton, GB--Southampton SO17 1BJ \and
Astronomical Institute, Academy of Sciences of the Czech Republic, CZ--25165 Ondrejov \and
CNR--Istituto di Astrofisica Spaziale e Fisica Cosmica, Sez. di Milano ``G. Ochialini'', Via Bassini 15, I--20133 Milano, Italy \and
Service d'Astrophysique, Centre d'Etudes de Saclay, F--91190 Gif-sur-Yvette Cedex, France \and
Laboratoire d'Astrophysique de Marseille, BP 8,  13376 Marseille cedex 12, France \and
Istituto di Astrofisica Spaziale, Area di Ricerca di Tor Vergata, Via del Fosso del Cavaliere, I--00133 Roma, Italy \and
Max-Planck-Institut f\"ur extraterrestrische Physik, Giessenbachstrasse , D--85748 Garching, Germany \and
%Institut d'Astrophysique et de G\'eophysique, Universit\'e de Li\`ege, 17, Allee du Six Aout, B--4000 Sart-Tilman Li\`ege, Belgium \and
Institut d'Astrophysique et de G\'eophysique, Univ. de Li\`ege, 17, Allee du Six Aout, B--4000 Sart-Tilman Li\`ege, Belgium \and
Danish Space Research Institute, Juliane Maries Vej 30, DK--2100 Copenhagen Ø, Denmark \and
Laboratory for High Energy Astrophysics, NASA's 
Goddard Space Flight Center, Code 660, Greenbelt, MD 20771 USA  \and
ESA-ESTEC, RSSD, P.O. Box 299, NL-2200 AG Noordwijk, The Netherlands \and
N. Copernicus Astronomical Ctr., Bartycka 18, PL--00716 Warsaw, Poland
}
\offprints{T. Courvoisier (ISDC)}
\mail{Thierry.Courvoisier@obs.unige.ch}
\date{Received /Accepted }
\abstract{
The {\it INTEGRAL} Science Data Centre (ISDC) provides the {\it INTEGRAL} data and means to analyse them to the scientific community. The ISDC runs a gamma ray burst alert system that provides the position of gamma ray bursts on the sky within seconds to the community. It operates a quick-look analysis of the data within few hours that detects new and unexpected sources as well as it monitors the instruments. The ISDC processes the data through a standard analysis the results of which are provided to the observers together with their data.
}
\authorrunning{T. J.-L. Courvoisier, R. Walter et al.}
\maketitle

\section{Introduction}
%\label{sec:introduction}

{\it INTEGRAL} is observing the gamma ray sky up to $10$\,{\rm MeV} with two instruments, the spectrometer SPI (\cite{vedrenneetal2003}) and the imager IBIS (\cite{ubertinietal2003}). The wavelength coverage is extended to X-rays ($3$\,{\rm keV}) by the X-ray monitor JEM-X (\cite{lundetal2003}) and to the optical domain by the OMC camera (\cite{mashesseetal2003}).

{\it INTEGRAL} has been conceived from the origin as an observatory type mission in which most of the observation time is available to the astronomical community at large, irrespective of their contribution to the building of the instruments.

Most of the {\it INTEGRAL} observations are long (several 100\,000\,s) and organised in many individual pointings of few 1000\,s (see below). The resulting data sets are large ($\simeq\,10$\,{\rm Gbytes} per day of observation) and complex as they need to refer to auxiliary data that describe the pointing and conditions in which each of the pointings was performed. In addition the data  for the three high-energy instruments are obtained from coded-mask optical systems and therefore require considerable processing before an image of the sky and/or the spectral energy distribution of a source can be calculated.

Well aware of the complexity of the data analysis and of the need to bring the data to the community in a form in which scientific results can be obtained it was decided very early on to develop a centre in which the data would be received, analysed, archived and distributed. The {\it INTEGRAL} Science Data Centre (ISDC) was conceived to meet these general goals. It grew from a consortium of European and US institutes who responded to a call by ESA to develop and operate the facility.

We will describe here the {\it INTEGRAL} observation strategy, the flow of data from {\it INTEGRAL} to the users, the main steps of the processing and the products that can be obtained from the ISDC to exploit the data.

\section{{\it INTEGRAL} observation strategy}

The {\it INTEGRAL} observation strategy is dictated by the coded-mask system. In turn this strategy drives the structure given to the {\it INTEGRAL}  data, it is therefore summarised here. 

The high energy instruments on board {\it INTEGRAL} are based on  coded-mask optics.  One of the properties of these systems, in particular in the case of the spectrometer SPI (\cite{vedrenneetal2003}), is that the number of detectors (and therefore of measurements during a single pointing) is less than the number of pixels on the sky in the field of view. In order to be able to exploit the angular resolution and the whole field of view of the spectrometer despite this property, but also to increase the number of background detector measurements during the observation, the {\it INTEGRAL} observations are split into many pointings separated by 2 degrees arranged in either a hexagonal pattern of 7 points around the main target of the observation or in a square of 25 pointings also centered on the main target of the observation (Fig. 1; \cite{jensenetal2003}). The dwell time on each pointing can vary based on the total time to be spent on a given dither pattern, it is typically 2000\,s. 

\begin{figure}
\includegraphics[width=0.5\textwidth]{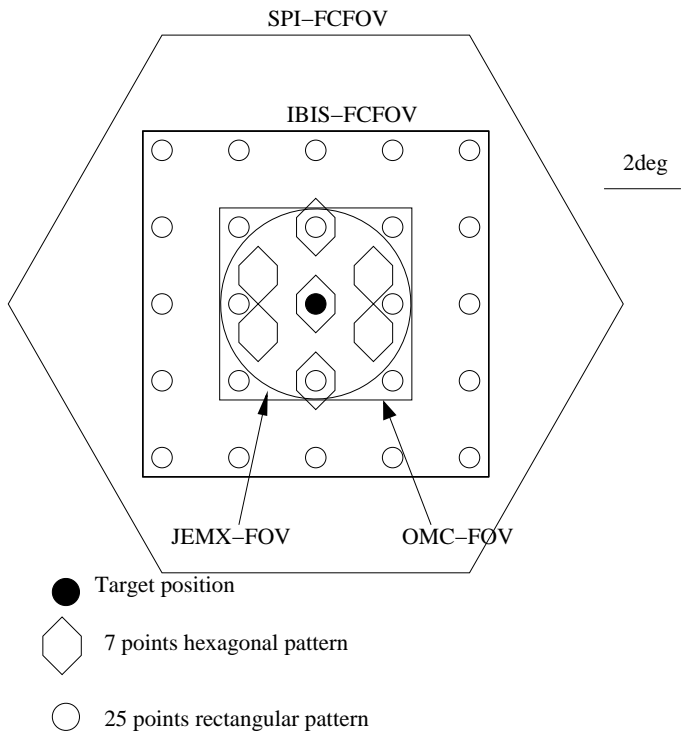}
%\hfill
%\parbox[b]{55mm}{
\caption{\label{fig:geometry}
 Hexagonal and 25 point dither patterns  }
%}
\end{figure}

The observations  of the Galactic plane or of the central regions of the Galaxy that correspond to the surveys of the core program (\cite{winkleretal2003})   are also organised in pointings of typically 2200\,s, albeit on patterns that are adapted to the surveys rather than in the hexagons or squares described above.

It follows that {\it INTEGRAL} data are divided in a (often large) number of  data sets, each associated with a single pointing or slew. Single pointings and slews are called science windows, they are the most important unit in the data processing and analysis.

\section{The {\it INTEGRAL} data Flow}

The scheduling of the observations is organised by the {\it INTEGRAL} Science Operations Centre (ISOC) at ESTEC (\cite{muchetal2003}). The commands to the satellite are generated by the European Space Operations Center  (ESOC) in Darmstadt (Germany).

The {\it INTEGRAL} data are received at the rate of 120\,kbits/s at one of two ground stations, Redu  in Belgium and Goldstone in California. The data are then routed to the ESOC  where the technical information is decoded and the status of the spacecraft and instruments checked. The data, technical and scientific, are then sent still in real time to the ISDC which is located in Versoix near Geneva. 

The data arrive at the ISDC with a delay of typically few seconds following their receipt at the ground station. The fraction of data that arrive in real time is larger than 95\%.

In order to correct for some problems that occur in the real time data transmission, the telemetry is consolidated few days after the event at ESOC. The consolidated data, that correspond to the most complete data set that can be obtained, are sent on compact disks to the ISDC approximately 2 weeks after the observation. 

\section{The ISDC processing}

The two data streams, the near real time data that arrive few seconds after reception on the ground  and the  consolidated data that arrive some two weeks later, are both processed at the ISDC. 

The near real time data are screened for gamma ray bursts  in the IBAS ({\it INTEGRAL} Burst Alert System, \cite{mereghettietal2003}) as they arrive. The bursts that occur in the field of view of the instruments are detected and their position on the sky calculated using the best attitude information available at the time of the burst within few seconds. These positions are sent in near real time to the community. Up to June  2003, 6 burst positions have been derived. IBAS has issued  one position within 30\,s of the burst (\cite{mereghettietal2003}). The IBAS system also issues alerts when bursts are observed in the anticoincidence shield of the SPI instrument. These bursts, although with no position on the sky from the {\it INTEGRAL} data, can be used in conjunction with other satellites of the interplanetary network, to measure their direction (\cite{vonkienlinetal2003}). There is approximately one burst per day in the anticoincidence shield data and one per month in the field of view of IBIS.

The near real time data are processed in a number of steps and merged with  data describing the planned operations (planning data) automatically. This processing generates a view of the instrument and spacecraft status, provides detector images and deconvolved sky images on a pointing by pointing basis. These data and products are  ready approximately 2 hours after their reception. Source detection algorithms are applied to the deconvolved data and the resulting source fluxes are  automatically compared with the content of a catalogue that describes the available knowledge of the field observed (\cite{ebisawaetal2003}). This quick look analysis (QLA) is able to detect and bring to the attention of ISDC scientists all sources brighter than $50-100$\,{\rm mCrab} in well defined energy bands and to report on any source that is the object of a target of opportunity proposal.

The consolidated data are processed automatically once received from ESOC through a standard analysis that is very similar to the quick-look analysis (see below).  One step is, however, added in which data from all pointings of an observation are combined together. The results of the analysis are then archived together  with the consolidated data in the {\it INTEGRAL} archive where they can be accessed when they become available according to the data rights in force  (\url{http://isdc.unige.ch}). The data and results of the standard analysis are finally distributed to the observer.

In addition to the automatic processing of the near real time and consolidated data, users can access either form of the data (near real time or consolidated) to perform analysis interactively through a set of programs called the Offline Science Analysis (OSA). This analysis is built using the same elements as the quick-look and standard analysis and generates the same products, albeit customised to the needs of the user through an interactive system. When used at ISDC on the near real time data, this analysis allows scientists to describe the properties of newly discovered sources within few hours. This has been at the origin of most of the communications to the community in the first months of the mission. The OSA software is distributed by the ISDC to the community. It can be obtained through the web  page of the ISDC. The software is distributed together with a description of the instruments in a format matched to the software, a set of test data and with an extensive documentation that includes not only user manuals and a scientific validation report for the instrument analysis but also a description of the known limitations and issues  of the version being distributed. A catalogue that describes known high energy sources (\cite{ebisawaetal2003}) is also provided. Regular releases of the complete package are taking place as the level of the software and instrument descriptions improve.

\begin{figure}[tb]
\includegraphics[width=0.47\textwidth]{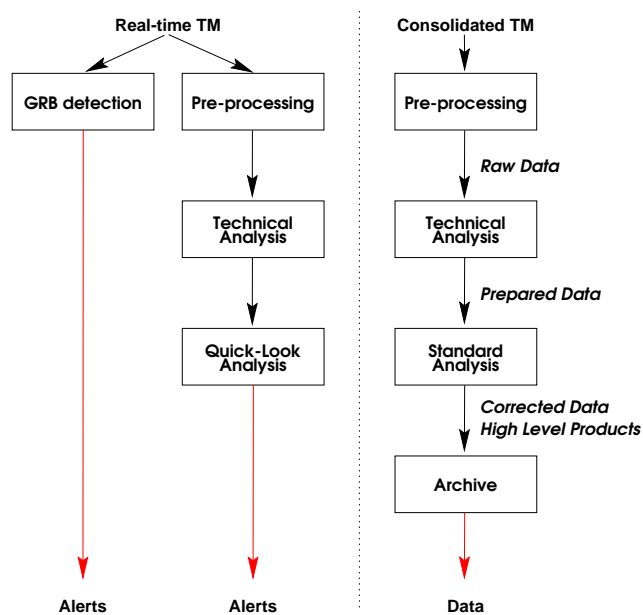}
%\hfill
%\parbox[b]{55mm}{
\caption{\label{fig:dataflow}
 The flow of data in the ISDC system  }
%}
\end{figure}

\section{{\it INTEGRAL} data}

The telemetry data ($120$\,{\rm kbits/s}) are received at the ISDC where a number of products combining instrument and spacecraft status, housekeeping and planning data are generated to provide the information required for the scientific exploitation in a convenient form.

\subsection{Raw data}

The raw data have the same information content as the original telemetry. These data are produced by the  pre-processing (Fig. 2). They are formatted and organised so that all data associated with a science window are grouped together, even if the telemetry is received at different times as is the case for histograms accumulated on board. 

{\it INTEGRAL} spacecraft and instruments generate many different kinds of data depending on the mode of the instruments. Consequently, pre-processing generates some 120 different types of data. 

Some types of data are not strictly associated with science windows and are associated with whole revolutions, they are processed and organised accordingly. These data include for example  descriptions of the status of the pixel array of the ISGRI detector or parameters that describe the on-board processing.  

\subsection{Prepared data}

The first step in the analysis is the conversion of all time formats to  a common  unit. This information is added to the raw data to form the prepared data (Fig.~2). The central on board time of the spacecraft is broadcast to the instruments which then use it individually. In addition SPI uses a further clock also synchronised with the spacecraft. All timing information is uniformised at this step to simplify all subsequent treatment of time data. On board time is used in all the ISDC processing to avoid any loss of information or precision. Conversion to standard barycentric time is made in the last steps of the analysis if and when necessary. The precision of the timing  is described in \cite{walteretal2003}.

In addition the technical analysis  performs several  tasks like an analysis of the instrument configuration at all times  that extracts lists of parameters to be used in the scientific processing. These parameters include descriptions of the pixels and parameters of the on board software that influence the data taken.  Instrument gains are also calculated and monitored.

This analysis step includes also a number of operations that  verify the status of the instruments and spacecraft and display status parameters for the operators.  It is thus possible to compute the fraction of time that is actually spent performing observations in nominal conditions while the spacecraft is stably pointing, an information that is provided to the mission planners who monitor the successful completion of the observation program. Operators and scientists are alerted whenever non  nominal conditions that may require actions are met. This analysis step also confronts the actual state of the instruments and spacecraft with the expected status derived from planning data.

\subsection{Scientific analysis}

The next step is the scientific analysis. This is done in the near real time environment by the  Quick Look Analysis (QLA), in the consolidated data stream automatically on all observations by the Standard Analysis (SA) and interactively by the users on any set of data by the Offline Science Analysis (OSA). This analysis  performs, irrespective of the environment where it runs, a number of corrections that are derived from the  calibrations files in the archive or obtained from the previous steps in the processing. Appropriate calibration data are selected from the status of the instruments, gains previously derived are applied and corrections related e.g. to the JEM-X anodes and to the status of the individual IBIS pixels are applied. The dead time for each detector is computed and used in the calculation of the fluxes. 

The time intervals during which the instruments were in conditions that are required for the analysis are computed. Only the corresponding data are used in the subsequent steps and the time intervals are used to calculate event rates. When the analysis is performed manually, the parameters defining these intervals can be set by the users. Pre-defined sets are used in the automatic processing like the quick-look analysis or the standard analysis.

High level products like deconvolved images, light curves and spectra are finally derived. This is done in standard forms for the quick-look and standard analysis and may be tuned by the user when the tools are used off-line. The algorithms used in this step are described elsewhere in this issue (\cite{goldwurmetal2003,mashesseetal2003,skinnerandconnell2003,strong2003}, \cite{westergaardetal2003}). Corrections that take the  alignment of the instruments compared to the reference attitude of the spacecraft given by the star trackers are applied. The  precision of the celestial coordinates provided for the sources is given in \cite{walteretal2003}.

All calibrations and instrument characteristics that are needed in the different steps of the analysis are stored in a set of files. For each file the epoch of its creation as well as the epochs for which it is valid are provided. The files are generated as knowledge of the instruments increases and as their charactistics change with time. They are accessed through the ISDC archive. They are organised so that the analysis software can access them without manual intervention.

All the products that are generated as the analysis proceeds are indexed in such a way that they are associated with the relevant original data without the need to duplicate these unnecessarily. The total amount of data including the raw telemetry generated per revolution is about $10$\,{\rm Gbytes}. It is organised in more than 10\,000 files in a well defined tree structure that is known to the software and largely transparent for the user.

In the near real time environment, the data and products are temporarily saved to be available to scientists responsible for following the observations and monitoring the instruments and to the observers on site.

In the consolidated data processing chain the data and products are archived and distributed to the users. The archive can be consulted from the network and public data retrieved. Data and products are also distributed to the observers by the ISDC.

The ISDC system has been designed to work for all instruments of {\it INTEGRAL}. The system is structured in a very modular architecture that allows a very easy adaptation to other instruments and missions. It is thus now being adapted to process data from one of the instruments of ESA's {\it Planck} mission. Further uses of the system can be envisaged in the future.

\section{The ISDC operations}

The ISDC system is operated around the clock in a dedicated set of machines organised as the operational network. Both near real time and consolidated data are processed and analysed in this environement.  Data in the operational network are not accessible from outside. The archive being populated from the operational network and accessible from outside according to data access rules is the only exception. Scientists and engineers analysing their own  data at the ISDC work on another set of computers organised as the office network. The office network (including the archive) has excellent connections to the external world (100\,{\rm megabits/s} in June 2003), while  the operational network is almost completly isolated. The ISDC work on the operational network is performed by a team of operators and scientists during office hours every day of the year while the development, maintenance, further data analysis and other tasks are performed by the remaining staff on the office network.

Variability is one of the main characteristics of most gamma-ray sources. In order to exploit this property to gain physical insights on the sources it is necessary to  react fast to unexpected events. The ISDC therefore reacts to gamma ray bursts in a matter of seconds and to the discovery of new sources in a matter of hours. To achieve this despite the fact that staff are present for only about a third of the time on site, the IBAS system and the near real time processing chain issue alerts automatically when gamma ray bursts occur,  when new sources are detected or when large amplitude variations with respect to expected source fluxe are measured. Alerts are also generated when major problems are found in a process or when communications in the ground segment are interrupted. 

These alerts are monitored continuously, the most important of them issue telephone messages to some of the staff $24$\,{\rm h/day}, allowing them to react in few hours at any time. This enables the ISDC to inform the community  of important events in few hours and thus it makes it possible for   ESA  and other agencies to perform target of opportunity observations with {\it INTEGRAL, XMM-Newton} or with other instruments in a short time. IBAS  sends a set of messages to the community within seconds of the detection of a gamma ray burst.

The near real time data are available at ISDC for analysis beyond the quick-look analysis approximately 2 hours after the observation was made. These data may be used by the observers present at ISDC while their data are being obtained. This facility has allowed to communicate the main characteristics of newly discovered sources within a short time to the community.

The ISDC  responds to enquiries of users related to data, analysis and software issues  through a helpdesk accessible through the ISDC web page. This is operated together with the ISOC so that users do not need to know whether their questions are best dealt with in ISOC or ISDC. Further forms of support to the community offered by ISDC staff include the possibility to visit the centre and the issue of an electronic newsletter. Data analysis workshops will also be organised regularly during the mission.

\section{The ISDC organisation}

The ISDC was established in 1995 after ESA had accepted the proposal put forward to them by a consortium of 12 institutes. The ISDC is attached to the Geneva Observatory, itself the astronomy department of the university of Geneva. This gives the possibility to accept a number of PhD students, of which there are 5 in Summer 2003. The staff increased steadily from very few (2) in 1995 to some 40 in June 2003 most of them working in Ecogia, a rural settlement near Versoix, some 8km from Geneva. The staff are mainly astrophysicists and software engineers, aproximatly half of them are under contract with co-investigators to work at the ISDC. Some instrument teams also keep staff located at ISDC to benfit from a close link.

The ISDC system has been designed, developed and implemented by the growing staff in close collaboration with the teams that developed the instruments and who contributed large parts of the instrument specific analysis software.  

\section{Conclusion}

The ISDC has been designed to be the interface between the {\it INTEGRAL} data and the scientific community. It aims at providing to the users of {\it INTEGRAL} the data, software and infrastructure that are needed to process these complex data in a way that makes it possible for general users to fully exploit the scientific potential of the mission. 

Since launch the ISDC system has been running without problems, all subsystems have been activated successfully. This resulted  in the timely publication of the first important {\it INTEGRAL} results on gamma-ray bursts and new  sources.

The archive is regularly populated and the data are sent to the users. No backlog has been generated over the first 8 months of the mission. 

The ISDC maintains a web page through which data, software, its newletter, general information and links to other sites are provided. This is accessible through \url{http://isdc.unige.ch}.

\begin{acknowledgements}
We are very indebted to all staff who have contributed in the course of the years to the development of the ISDC, be they staff of the ISDC, of the institutes who contributed to the instruments  or of ESA.

The ISDC is attached to the Geneva Observatory. It is funded by national funding agencies and ESA through the Principal and Co- investigators who form the ISDC consortium (T.J.-L. Courvoisier, 
R. Walter,
A.J. Dean,
R. Hudec,
S. Mereghetti,
T. Montmerle,
A. Preite Martinez,
R. Staubert,
A.W. Strong,
J.-P. Swings,
N.J. Westergaard,
N. White,
A.A. Zdziarski).  The {\it INTEGRAL} instrument teams have contributed a large fraction of the instrument specific software.

The ISDC development benefited from several visits made at similar centers in Europe and the US. 
\end{acknowledgements}

\end{document}